\documentclass{ws-ijmpd}

\begin{document}
\newcommand{\beq}[1]{\begin{equation}\label{#1}}
 \newcommand{\eeq}{\end{equation}}
 \newcommand{\bea}{\begin{eqnarray}}
 \newcommand{\eea}{\end{eqnarray}}
 \def\disp{\displaystyle}
\def\l{\left}
\def\r{\right}

%\markboth{Authors' Names}
%{Instructions for Typing Manuscripts (Paper's Title)}

\catchline{}{}{}{}{}
%
%%%%%%%%%%%%%%%%%%%%%%%%%%%%%%%%%%%%%%%%%%%%%%%%%%%%%%%%%%%%%%%%%%%%

\title{Calibration of GRB Luminosity Relations with Cosmography }

\author{He Gao}

\address{Department of Astronomy, Beijing Normal   University, Beijing 100875, China\\
Department of Physics and Astronomy, University of Nevada Las Vegas,
4505 Maryland Parkway, Box 454002, Las Vegas, NV 89154-4002,
USA\\gaohe@physics.unlv.edu}

\author{Nan Liang}

\address{Center for High Energy Physics, Peking University, Beijing
100871,  China\\Department of Astronomy, Beijing Normal   University, Beijing 100875, China\\
liangn@bnu.edu.cn}

\author{Zong-Hong Zhu}

\address{Department of Astronomy, Beijing Normal   University, Beijing 100875, China\\
zhuzh@bnu.edu.cn}

\maketitle

%\begin{history}
%\received{Day Month Year}
%\revised{Day Month Year}
%\comby{Managing Editor}
%\end{history}

\begin{abstract}
For the use of Gamma-Ray Bursts (GRBs) to probe cosmology in a
cosmology-independent way, a new method has been proposed to obtain
luminosity distances of GRBs by interpolating directly from the
Hubble diagram of SNe Ia, and then calibrating GRB relations at high
redshift. In this paper, following the basic assumption in the
interpolation method that objects at the same redshift should have
the same luminosity distance, we propose another approach to
calibrate GRB luminosity relations with cosmographic fitting
directly from SN Ia data. In cosmography, there is a well-known
fitting formula which can reflect the Hubble relation between
luminosity distance and redshift with cosmographic parameters which
can be fitted from observation data. Using the Cosmographic fitting
results from the Union set of SNe Ia, we calibrate five GRB
relations using GRB sample at $z\leq1.4$ and deduce distance moduli
of GRBs at $1.4< z \leq 6.6$ by generalizing above calibrated
relations at high redshift. Finally, we constrain the dark energy
parameterization models of the Chevallier-Polarski-Linder (CPL)
model, the Jassal-Bagla-Padmanabhan (JBP) model and the Alam model
with GRB data at high redshift, as well as with the Cosmic Microwave
Background radiation (CMB) and the baryonic acoustic oscillation
(BAO) observations, and we find the $\Lambda$CDM model is consistent
with the current data in 1-$\sigma$ confidence region.
\end{abstract}


\section{Introduction}
Since an intrinsic relation between the peak luminosity and the
shape of the light curve of SNe Ia has been
found,\cite{Phillips1993} SNe Ia has now been taken as near-ideal
standard candles for measuring the geometry and dynamics of the
universe. However, the maximum redshift of the SNe Ia which we can
currently use is only about 1.7. On the other hand, the redshift of
the last scattering surface of the cosmic microwave background (CMB)
is at $z=1091.3$.\cite{Komatsu2010}

Recently, Gamma-Ray Bursts (GRBs) were proposed to be a
complementary probe to SNe Ia and CMB to explore the early universe.
As the most intense explosions observed in the universe so far, GRBs
are likely to occur in high-redshift range up to at least
$z=8.2$.\cite{Tanvir2009,Salvaterra2009} Moreover, there are several
luminosity relations of GRBs between the spectral and temporal
properties which have been extensive discussed, such as the
isotropic energy ($E_{\rm iso}$) - peak spectral energy ($E_{\rm
peak}$) relation,\cite{Amati2002} the luminosity ($L$) - spectral
lag ($\tau_{\rm lag}$) relation,\cite{Norris2000} the $L$ -
variability ($V$) relation,\cite{Fenimore2000,Riechart2001} the $L$
- $E_{\rm peak}$ relation,\cite{Schaefer2003a,Yonetoku2004} the $L$
- minimum rise time ($\tau_{\rm RT}$) relation,\cite{Schaefer2002}
and the collimation-corrected energy ($E_{\gamma}$) - $E_{\rm peak}$
relation;\cite{Ghirlanda2004a} as well as several multiple relations
such as the $E_{\rm iso}$ - $E_{\rm peak}$ - $t_{\rm b}$
relation,\cite{Liang2005} where $t_{\rm b}$ is the break time of the
optical afterglow light curves; the $L$ - $E_{\rm peak}$ -
$T_{0.45}$ relation,\cite{Firmani2006} where $T_{0.45}$ is the
rest-frame ``high-signal'' timescale; and the $L$ - $E_{\rm peak}$ -
$\tau_{\rm lag}$ (or $\tau_{\rm RT}$) relation.\cite{Yu2009}

Many authors have made use of GRB luminosity indicators as standard
candles at very high redshift beyond SNe Ia redshift range for
cosmological
research.\cite{Schaefer2003b,Takahashi2003,Bloom2003,Dai2004,Ghirlanda2004a,Ghirlanda2004b,Friedman2005,Firmani2006,Firmani2007,Liang2005,Xu2005,Wang2006,Bertolami2006,Ghirlanda2006,Schaefer2007,Wright2007,Wang2007,Basilakos2008,Cuesta2008a,Cuesta2008b,Daly2008,Qi2008a,Qi2008b,Qi2009}
Due to the lack of the sample at low redshift which are cosmology
independent, to calibrate the empirical GRB relations, one usually
needs to assume a particular cosmological model with certain model
parameters as a $priori$. As a result, the so-called circularity
problem could prevent the direct use of GRBs for
cosmology.\cite{Ghirlanda2006} Many of works treat the circularity
problem with statistical approach which carried out a simultaneous
fit of the parameters in the calibration curves and the
cosmology.\cite{Schaefer2003b,Firmani2005,Li2008,Amati2008,WangY2008}
However, the circularity problem can not be circumvented completely
by means of the statistical approaches for an input cosmological
model is still required in doing the joint fitting.

More recently, Liang \textit{et al.} proposed a new method to
calibrate GRB luminosity relations in a cosmological
model-independent way.\cite{Liang2008a} The motivation of this
calibration  method is that objects at the same redshift should have
the same luminosity distance in any cosmology. Thus the luminosity
distance of a GRB at a given redshift can be obtained by
interpolating directly from the Hubble diagram of SNe Ia, therefore
GRB relations can be calibrated without assuming a particular
cosmological model and the Hubble diagram of GRBs  has been
constructed. Following this cosmology-independent GRB calibration
directly from SNe Ia, the derived GRB Hubble diagram  can be used to
constrain cosmological models at high redshift avoiding circularity
problem
\cite{Liang2008b,Capozziello2008,Izzo2009,Wei2009a,Wei2009b,Wang2009,Liang2010,WLiang2010}.
Capozziello \& Izzo firstly used two GRB relations calibrated with
the so-called Liang method to derive the related cosmography
parameters which related to the derivatives of the scale
factor.\cite{Capozziello2008} Liang \textit{et al.} combined the
updated distance moduli of GRBs obtained by the interpolating method
with the joint data to find the contribution of GRBs to the joint
cosmological constraints in the confidence regions of cosmological
parameters, and reconstructed the acceleration history of the
universe with the distance moduli of SNe Ia and
GRBs.\cite{Liang2010} On the other hand, besides the interpolation
method, the luminosity distance of a GRB  can also be obtained
directly from SNe Ia data by other mathematical approach. Liang \&
Zhang has proposed another approach to calibrate GRB relations by
using an iterative procedure which is a non-parametric method in a
model independent manner to reconstruct the luminosity distance at
any redshift from SNe Ia.\cite{Liang2008b} Similar to the
interpolation method, Cardone \textit{et al.} constructed an updated
GRBs Hubble diagram calibrated by local regression from SNe
Ia.\cite{Cardone2009} Kodama \textit{et al.} has proposed that the
$L-E_{\rm peak}$ relation can be calibrated with one empirical
formula fitted from the luminosity distance of SNe
Ia.\cite{Kodama2008} However, according to the formula fitting
approach, various possible formula can be fitted from the SNe Ia
data which could give different calibration results of GRB
relations. As the cosmological constraints from GRBs are sensitive
to GRBs calibration results, and the fitting procedure depends
seriously on the choice of the formula, the reliability of this
method should be tested carefully. In other words, we should find
one certain formula which is totally independent of any cosmological
models and could accurately evaluate the Hubble relation.

In Cosmography,\cite{Visser2004} there is a well-known formula
reflecting the Hubble relation between luminosity distance and
redshift which can be extracted directly from basic cosmological
principles and observation data, with cosmography parameters (the
deceleration, jerk and snap parameters: $q$, $j$, and $s$)  which
are only related to the derivatives of the scale factor without any
priori assumption on the underlying cosmological model. Recently,
several authors have already used the cosmographic parameters
fitting from SNe Ia and/or GRBs dataset to constrain cosmological
parameters.\cite{Cattoen2007,Cattoen2008,Capozziello2008,Vitagliano2010}
If viewing this point from another angle, the cosmographic formula
can be considered as a perfect fitting function to calibrate the GRB
relations using SNe Ia data, as long as we take the same assumption
that objects at the same redshift should have the same luminosity
distance in any cosmology. In this paper, instead of the
interpolation method using in Ref.~\refcite{Liang2008a}, we propose
another new approach to calibrate GRB luminosity relations with
cosmographic fitting from SNe Ia data. The structure of this paper
is arranged as follows. In section 2 we give a brief review of the
cosmographic Hubble relation between luminosity distance and
redshift. In section 3, we calibrate five GRB luminosity relations
with cosmographic fitting results from SNe Ia data. In section 4, we
construct the Hubble diagram of GRBs obtained by using the
cosmographic methods and  constrain the dark energy parameterization
models of the Chevallier-Polarski-Linder (CPL)
model,\cite{Chevallier2001,Linder2003} the Jassal-Bagla-Padmanabhan
(JBP) model\cite{Jassal2004} and the Alam model\cite{Alam2003} with
GRB data at high redshift, as well as with the Cosmic Microwave
Background radiation (CMB) and the baryonic acoustic oscillation
(BAO) observations. Conclusions and discussions are given in section
5.

\section{Cosmographic Hubble relation between luminosity distance and redshift}

A completely new cosmology branch -- cosmography, in which framework
cosmology is pure kinematics and completely independent of the
underlying dynamics governing the evolution of the universe has been
introduced since 1970s.\cite{Wenberg1972} The only assumption is the
basic symmetry principles (the cosmological principle) that the
universe can be described by the Friedmann-Robertson-Walker metric.
By means of a Taylor series expansion of the luminosity distance,
Visser gave a formulation that the luminosity distance can be
expressed as a power series in the redshift up to the forth
order:\cite{Visser2004}
 \bea
d_{L}(z)=&&cH_0^{-1}\l\{z+\frac{1}{2}(1-q_0)z^2-\frac{1}{6}\l(1-q_0-3q_0^2+j_0+
\frac{kc^2}{H_0^2a^2(t_0)}\r)z^3+\r.\nonumber \\ && \nonumber \\
&&\hspace{-1cm}\l.+\frac{1}{24}\l[2-2q_0-15q_0^2-15q_0^3+5j_0+10q_0j_0+
s_0+\frac{2kc^2(1+3q_0)}{H_0^2a^2(t_0)}\r]z^4+...\r\}\, ,\eea where
the coefficients of the expansion are the so-called cosmographic
parameters, Hubble parameters ($H\equiv{\dot{a}(t)}/{a(t)}$),
deceleration parameters ($q\equiv-H^{-2}{\ddot{a}(t)}/{a(t)}$), jerk
parameters ($j\equiv H^{-3}{a^{(3)}(t)}/{a(t)}$), and snap
parameters ($s\equiv H^{-4}{a^{(4)}(t)}/{a(t)}$), which related to
the scale factor $a(t)$ and its higher order derivatives (the
subscript ``0'' indicates the present value of the parameters).
Obviously pure cosmography by itself will not predict anything about
the scale factor $a(t)$, we have to turn to the observational data
such as SNe Ia to infer the history of the scale factor $a(t)$ and
some important information about expanding history of our universe.

In order to avoid problems with the convergence of the series for
the highest redshift objects as well as to control properly the
approximation induced by truncations of the expansions, Cattoen \&
Visser  pointed out that it is useful to recast $d_L$ as a function
of an improved parameter $y=z/(1+z)$ and constrained the
cosmographic parameters using SNe Ia data.\cite{Cattoen2007} In such
a way, being $z\in(0,\infty)$ mapped into $y\in(0,1)$, the
luminosity distance at the fourth order in the $y$ - parameter
becomes:
\begin{eqnarray}
d_L(y)=&&\frac{c}{H_0}\left\{y-\frac{1}{2}(q_0-3)y^2+\frac{1}{6}\left[12-5q_0+3q^2_0-(j_0+\Omega_0)\right]y^3+\frac{1}{24}\left[60-7j_0-\right.\right.\nonumber
  \\ &&\left.\left.-10\Omega_0-32q_0+10q_0j_0+6q_0\Omega_0+21q^2_0-15q^3_0+s_0\right]y^4+\mathcal{O}(y^5)\right\}~,
\end{eqnarray}
where $\Omega_0=1+kc^2/H_0^2a^2(t_0)$ is the total energy density.
For the flat universe, $\Omega_0=1$. The luminosity distance as the
logarithmic Hubble relations can be expressed as:
\begin{eqnarray}
\ln{\left[\frac{d_L(y)}{\rm
      Mpc}\right]}=\ln{y}+\ln{\left[\frac{c}{H_0}\right]}-\frac{1}{2}(q_0-3)y+\frac{1}{24}\left[21-4(j_0+\Omega_0)+q_0(9q_0-2)\right]y^2 \nonumber
      \\+\frac{1}{24}\left[15+4\Omega_0(q_0-1)+j_0(8q_0-1)-5q_0+2q^2_0-10q^3_0+s_0\right]y^3+\mathcal{O}(y^4)~,\label{eq:y-exp}
\end{eqnarray}
therefore the distance modulus can be given by
\begin{eqnarray}\mu(y)=25+\frac{5}{\ln{10}}\ln{\left[\frac{d_L(y)}{\rm
      Mpc}\right]}.\end{eqnarray}

Recently,  Vitagliano \textit{et al.} fitted two different
truncations (Cosmography I: without the third order term ($y^3$);
Cosmography II: with the third order term ($y^3$)) of the above
expansion  with the SNe Ia and  GRB datasets.\cite{Vitagliano2010}
With a flat universe ($\Omega_{0}=1$) prior, the fitting results
showed that the Union dataset\cite{Kowalski2008} gave more stringent
constraints on the parameters. For Cosmography I, the cosmographic
fitting results with the Union dataset are
\begin{equation}
q_0=-0.58\pm0.24~~,~~~j_0+\Omega_0=0.91\pm2.21\nonumber~,
\end{equation}
and for Cosmography II, the cosmographic fitting results with the
Union dataset are
\begin{equation}
q_0=-0.50\pm0.55~~,~~~j_0+\Omega_0=-0.26\pm9.0,~~~s_0=-4.13\pm129.79\nonumber~.
\end{equation}

It is noted that the Union dataset of 307 SNe  Ia  didn't include
the 90 SNe Ia data from CfA3\cite{Hicken2009} due to their extremely
low redshift ($z < 0.1$), which would not affect the calibrated
results for GRB luminosity relations at $ z \ge
0.17$.\cite{Liang2010} With the above cosmographic fitting results
from the Union dataset, the cosmographic Hubble relation can be
considered as a perfect function to deduce the distance moduli of
GRBs directly from SNe Ia data. In the next section, we will deduce
the distance moduli of GRBs and then calibrate the GRB luminosity
relations with Cosmography I and II respectively.

\section{The Calibration of the Luminosity Relations of Gamma-Ray Bursts}

We adopt the 69 GRBs provided in Ref.~\refcite{Schaefer2007} as our
sample for calibrating the GRB luminosity/energy relations. We first
deduce the distance moduli of GRBs at $z\leq1.4$ within our sample.
Then using these deduced distance moduli and the redshifts of
corresponding GRBs, we calibrate five GRB luminosity/energy
relations i.e., the $\tau_{\rm lag}$ - $L$ relation, the $V$ - $L$
relation, the $L$ - $E_{\rm peak}$ relation, the $E_{\gamma}$ -
$E_{\rm peak}$ relation and the $\tau_{\rm RT}$ - $L$ relation.
These luminosity relations of GRBs can be generally written in the
form
\begin{equation}
\log y=a+b\log x,
\end{equation}
where $a$ and $b$ are the intercept and slope of the relation
respectively, in addition we introduce c as the linear correlation
coefficient of the relation which will be calculated together with a
and b below; $y$ is the luminosity ($L/{\rm erg s^{-1}}$ or energy
$E_{\gamma}/\textrm{erg}$); $x$ is the GRB parameters measured in
the rest frame, e.g., $\tau_{\rm lag}(1+z)^{-1}/(0.1\rm \ s)$,
$V(1+z)/0.02$, $E_{\rm peak}(1+z)/(300\ \rm keV)$, $\tau_{\rm
RT}(1+z)^{-1}/\rm (0.1\ s)$, for the corresponding relations above.
For the $x$ values, we adopt the data from
Ref.~\refcite{Schaefer2007}; for the $y$ values, we drive them with
the adjusted luminosity distance of GRBs calculated with
cosmographic fitting method (Cosmography I and II). The isotropic
luminosity of a burst is calculated by
\begin{equation}
L = 4\pi d_{L}^2 P_{\rm bolo},
\end{equation}
where $P_{\rm bolo}$ is the bolometric flux of gamma-rays in the
burst and $d_{L}$ is the luminosity distance of the burst. The
isotropic energy released from a burst is given by
\begin{equation}
E_{\rm iso} = 4\pi d_{L}^2 S_{\rm bolo} (1+z)^{-1},
\end{equation}
where $S_{\rm bolo}$ is the bolometric fluence of gamma-rays in the
burst at redshift $z$. The total collimation-corrected energy is
then calculated by
\begin{equation}
E_{\gamma} = F_{\rm beam}E_{\rm iso},
\end{equation}
where the beaming factor, $F_{\rm beam}=(1-\cos\theta _{\rm jet})$;
and the value of the jet opening angle $\theta _{\rm{jet}}$ is
related to the jet break time ($t_{\rm b}$) and the isotropic energy
for an Earth-facing jet, $E_{\gamma ,\rm iso,52}=E_{\gamma ,\rm
iso}/10^{52} \rm erg$.  When calculating $E_{\gamma ,\rm iso}$, we
also use the cosmographic fitting method from SNe Ia to avoid the
circularity problem.

We determined the values of the intercept ($a$) and the slope ($b$)
with their 1-$\sigma$ uncertainties with the same regression method
(the bisector of the two ordinary least-squares)  used in
Ref.~\refcite{Schaefer2007,Liang2008a}. The calibrate results for
Cosmography I and II are summarized in Table 1. From Table 1, we
find that the calibration results obtained using two cosmographic
fitting (Cosmography I and II) are fully consistent with each other.
We also find that results obtained by the cosmographic methods
differ only slightly from, but still fully consistent with those
calibrated by using the interpolation method and the iterative
procedure with the same GRB sample (see details
inthere\cite{Liang2008a,Liang2008b}).

\begin{table}[ph]
\tbl{Calibration results (for $a$=intercept,  $b$=slope,
$c$=correlation coefficient) with their 1-$\sigma$ uncertainties,
for the five GRB luminosity/energy relations within the sample at
$z\le1.4$, using two cosmographic fitting results (Cosmography I and
II) directly from SNe Ia data.} {\begin{tabular}{lccccccccccc}
  \hline\hline
 & \multicolumn{3}{c|}{Cosmography I}

 & \multicolumn{3}{c}{Cosmography II} & \\
 \cline{2-4} \cline{5-7}Relation &
$a$&$b$&$c$  &   $a$&$b$&$c$& \\
 \hline
$\tau_{\rm lag}$-$L$                                    &52.13$\pm0.10$& -1.10$\pm0.13$&-0.89\ \ & 52.21$\pm0.11$& -1.13$\pm0.14$&-0.88\\
$V$-$L$                                                 &52.47$\pm0.13$&  2.02$\pm0.27$& 0.64\ \ & 52.54$\pm0.13$&  2.06$\pm0.27$& 0.65\\
$L$-$E_{\rm peak}$                                         &52.14$\pm0.09$&  1.64$\pm0.10$& 0.89\ \ & 52.21$\pm0.09$&  1.67$\pm0.10$& 0.89\\
$E_{\gamma}$-$E_{\rm peak}$                                &50.83$\pm0.06$&  1.87$\pm0.11$& 0.95\ \ & 50.88$\pm0.07$&  1.93$\pm0.10$& 0.95\\
$\tau_{\rm RT}$-$L$                                     &52.51$\pm0.11$& -1.29$\pm0.11$&-0.77\ \ & 52.58$\pm0.11$& -1.31$\pm0.12$&-0.77\\
\hline
\end{tabular}
}

\end{table}

%%==================== table 1 ==========================================================================
%\begin{table}[ph]
%\tbl{Comparison of acoustic for frequencies for piston-cylinder
%problem.}
%
%\begin{center}{\scriptsize
%{\begin{tabular}{lccccccccccc}
%  \hline\hline
% & \multicolumn{3}{c|}{Cosmography I}
%
% & \multicolumn{3}{c}{Cosmography II} & \\
% \cline{2-4} \cline{5-7}Relation &
%$a$&$b$&$c$  &   $a$&$b$&$c$& \\
% \hline
%$\tau_{\rm lag}$-$L$                                    &52.13$\pm0.10$& -1.10$\pm0.13$&-0.89\ \ & 52.21$\pm0.11$& -1.13$\pm0.14$&-0.88\\
%$V$-$L$                                                 &52.47$\pm0.13$&  2.02$\pm0.27$& 0.64\ \ & 52.54$\pm0.13$&  2.06$\pm0.27$& 0.65\\
%$L$-$E_{\rm peak}$                                         &52.14$\pm0.09$&  1.64$\pm0.10$& 0.89\ \ & 52.21$\pm0.09$&  1.67$\pm0.10$& 0.89\\
%$E_{\gamma}$-$E_{\rm peak}$                                &50.83$\pm0.06$&  1.87$\pm0.11$& 0.95\ \ & 50.88$\pm0.07$&  1.93$\pm0.10$& 0.95\\
%$\tau_{\rm RT}$-$L$                                     &52.51$\pm0.11$& -1.29$\pm0.11$&-0.77\ \ & 52.58$\pm0.11$& -1.31$\pm0.12$&-0.77\\
%\hline
%\end{tabular}}
%}
%\end{center}
% \caption{\label{tab1}Calibration results (for $a$=intercept,  $b$=slope,
%$c$=correlation coefficient) with their 1-$\sigma$ uncertainties,
%for the five GRB luminosity/energy relations within the sample at
%$z\le1.4$, using two cosmographic fitting results (Cosmography I and
%II) directly from SNe Ia data.}
%%\vspace{0.5cm}
%\end{table}

\section{The Hubble diagram of gamma-ray bursts}

%==================== Fig 1 ==========================================================================
 \begin{figure}[pb]
\begin{center}
\includegraphics[angle=0,scale=0.4]{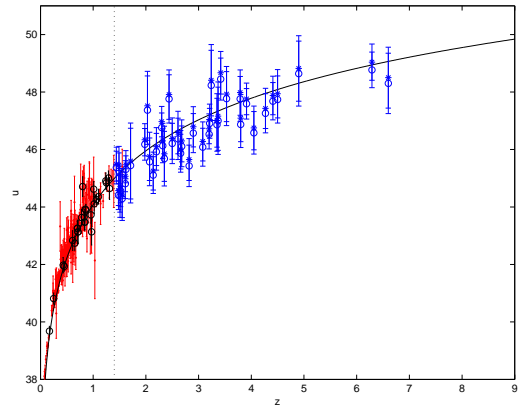}
\end{center}
\caption{Hubble Diagram of 307 SNe Ia (\textit{red dots}) and the 69
GRBs (\textit{circles}) obtained using the cosmographic method. The
27 GRBs at $z\le1.4$ are obtained by the cosmographic fitting from
SNe Ia data (\textit{black circles}), and the 42 GRBs at $z>1.4$
(\textit{blue circles}) are obtained with the five relations
calibrated with the sample at $z\le1.4$ using the cosmographic
method (\textit{blue circles}: Cosmographic I; \textit{blue stars}:
Cosmographic II). The curve is the theoretical distance modulus in
the concordance model ($w= -1$, $\Omega_{\rm M0}= 0.27$,
$\Omega_{\Lambda}= 0.73$), and the vertical dotted line represents
$z=1.4$.}
\end{figure}

With the cosmographic fitting results (Cosmography I or II), the
moduli for the 27 GRBs at $z\leq1.4$ can be directly obtained from
SNe data, therefore we calibrate GRB luminosity relations in a
completely cosmology-independent way. Furthermore, if assuming that
GRB luminosity relations do not evolve with redshift, we are able to
obtain the luminosity ($L$) or energy ($E_{\rm \gamma}$) of each
burst at high redshift ($z > 1.4$) by utilizing the calibrated
results from Cosmography. Consequently, the corresponding luminosity
distance ($d_L$) can be derived from Eq.(12) $\sim$ Eq.(14) and the
corresponding distance modulus can be calculated as
$\mu=5\log(d_L/{\rm Mpc})+25$. The uncertainty of the value of the
luminosity or energy deduced from a GRB relation is
\begin{equation}
\sigma_{\log y}^2 = \sigma_a^2 + (\sigma_b \log x)^2 + (0.4343 b
\sigma_x / x)^2 + \sigma_{\rm sys}^2,
\end{equation}
where $\sigma_{a}$, $\sigma_b$ and  $\sigma_{x}$ are 1-$\sigma$
uncertainty of the intercept, the slope and the GRB measurable
parameters, and $\sigma_{\rm sys}$ is the systematic error in the
fitting that accounts for the extra scatter of the luminosity
relations.\cite{Schaefer2007} Note that the uncertainty of modulus
for each luminosity indicator depends on whether $P_{\rm bolo}$ or
$S_{\rm bolo}$ is used:
\begin{equation}
\sigma_{\mu} = [(2.5 \sigma_{\log L})^2+(1.086 \sigma_{P_{\rm
bolo}}/P_{\rm bolo})^2]^{1/2},
\end{equation}
or
\begin{equation}
\sigma_{\mu} = [(2.5 \sigma_{\log E_{\gamma}})^2+(1.086
\sigma_{S_{\rm bolo}}/S_{\rm bolo})^2+(1.086 \sigma_{F_{\rm
beam}}/F_{\rm beam})^2]^{1/2}.
\end{equation}

For five luminosity indicators, each burst will have up to five
estimated distance moduli, we hence use the same method in
Ref.~\refcite{Schaefer2007} to obtain the best estimated $\mu$ for
each GRB which is the weighted average of all available distance
moduli:

\begin{equation}
\mu = (\sum_i \mu_{\rm i} / \sigma_{\mu_{\rm i}}^2)/(\sum_i
\sigma_{\mu_{\rm i}}^{-2}),
\end{equation}
with its uncertainty $ \sigma_{\mu} = (\sum_i \sigma_{\mu_{\rm
i}}^{-2})^{-1/2}$, where the summations run from 1 to 5  over the
five relations used in Ref.~\refcite{Schaefer2007} with available
data. Until now we have ultimately obtained the 42 GRB moduli at
$z>1.4$ by utilizing the five relations calibrated with the sample
at $z\le1.4$ using the fitting method. We have plotted the Hubble
diagram of 307 SNe Ia and the 69 GRBs obtained using the
cosmographic methods in Fig 1.

Constraints  from the GRB data can be obtained consequently by
fitting the distance moduli $\mu(z)$. The $\chi^2$ value of the
observed distance moduli can be calculated by
\begin{eqnarray}
\chi^2_{\mu}=\sum_{i=1}^{N}\frac{[\mu_{\rm{obs}}(z_i)-\mu(z_i)]^2}
{\sigma_{\mu,i}^2},
\end{eqnarray}
where $\mu _{\rm obs}(z_i)$ is the observed distance modulus for the
GRBs at redshift ~$z_i$~ with its error $\sigma_{\mu_{\rm i}}$;
$\mu(z_i)$ is the theoretical value of distance modulus from a dark
energy model. As mentioned above, high-redshift GRBs are rare
database for constraining cosmological parameters. Since those 42
GRBs' moduli ($1.4<z\le6.6$) calculated above are completely
cosmological model independent, we thus utilize this dataset
together with CMB and BAO observations to constrain specified
cosmological models.

For the CMB observation, we choose one shift parameter $R$ to limit
the model parameters. In a flat universe, it can be expressed as

\begin{eqnarray}
R=\Omega_M^{1/2}\int_0^{z_{ls}}{\frac{dz}{E(z)}},
\end{eqnarray}
where the last scattering redshift $z_{ls} = 1091.3$ from the 7-year
WMAP results, and the observational value
$R=1.725\pm0.018$\cite{Komatsu2010}. The $\chi_{\rm CMB}^2$ value is

\begin{eqnarray}
\chi_{\rm CMB}^2=\frac{(R-1.725)^2}{0.018^2}.
\end{eqnarray}

For the BAO observation, the size of baryon acoustic oscillation
peak can be used to constrain the cosmological
parameters.\cite{Blake2003,Seo2003,Dolney2006} This peak can be
denoted by a parameter A, which can be expressed
as\cite{Eisenstein2005}

\begin{eqnarray}
A=\Omega_M^{1/2}E(z_{\rm BAO})^{-1/3}\big[\frac{1}{z_{\rm
BAO}}\int_0^{z_{\rm BAO}}{\frac{dz}{E(z)}}\big]^{2/3},
\end{eqnarray}
 where $z_{\rm BAO} = 0.35$. The observational value is
$A=0.469(n_s/0.98)^{-0.35}\pm0.017$,\cite{Eisenstein2005} with the
scalar spectral index $n_s=0.963$ from WMAP7 data.\cite{Komatsu2010}
The $\chi_{\rm BAO}^2$ value is

\begin{eqnarray}
\chi_{\rm BAO}^2=\frac{[A-0.469(n_s/0.98)^{-0.35}]^2}{0.017^2}.
\end{eqnarray}

Here we combine these two probes with the GRBs dataset above by
multiplying the likelihood functions. The total $\chi^2$ value is

\begin{eqnarray}
\chi_{\rm total}^2=\chi_{\rm GRB}^2+\chi_{\rm CMB}^2+\chi_{\rm
BAO}^2.
\end{eqnarray}

%==================== table 2 ==========================================================================
\begin{table}[pb]
\tbl{The best-fit values of $w_0$ and $w_a$ (or $A_1$ and $A_2$)
with 1-$\sigma$ uncertainties, as well as the best-fit values of
$\Omega_M$, $H_0$, and $\chi_{\rm min}^2$, for the CPL model, JBP
model and Alam model, with GRB data (calibrated with Cosmography I
and II fitting from SN Ia data), as well as with CMB and BAO
observations. } {{\scriptsize
 \begin{tabular}{c|c|c|c|c|c|c} \hline\hline

& \multicolumn{2}{c|}{CPL Model} & \multicolumn{2}{c|}{JBP Model}
& \multicolumn{2}{c}{Alam Model}\\
 \cline{2-3} \cline{4-5}\cline{6-7} & Cosmography I& Cosmography II  & Cosmography I&Cosmography II & Cosmography I&Cosmography II\\
      \hline
  $w_0(A_1)$    \ \ & \ \ $-0.79^{+0.38}_{-0.34}$\ \     & \ \ $-0.91^{+0.38}_{-0.32}$\ \            & \ \ $-0.74^{+0.53}_{-0.51}$\ \      & \ \ $-1.37^{+0.52}_{-0.51}$\ \     & \ \ $0.60^{+0.96}_{-0.97}$\ \      & \ \ $-0.39^{+0.90}_{-0.87}  $\ \ \\
  $w_a(A_2)$   \ \ & \ \ $-0.36^{+1.15}_{-1.7}$\ \     & \ \ $-0.06^{+1.06}_{-1.66}$\ \            & \ \ $-0.93^{+3.17}_{-3.53}$ \ \    & \ \ $2.02^{+3.16}_{-3.48}$\ \     & \ \ $-0.11^{+0.26}_{-0.25}$\ \      & \ \ $0.10^{+0.24}_{-0.24}  $\ \ \\
      \hline
 $\Omega_M$    \ \ & \ \ $0.29$\ \                  & \ \ $0.28$\ \                            & \ \ $0.29$\ \                      & \ \ $0.25$\ \                        & \ \ $0.29$\ \                   & \ \ $0.26$\ \ \\
  $H_0$   \ \      & \ \ $80$\ \                      & \ \ $75$\ \                            & \ \ $80$ \ \                         & \ \ $79$\ \                      & \ \ $79$\ \                       & \ \ $78$\ \ \\
$\chi_{\rm min}^{2}$   \ \  & \ \ $28.03$\ \          & \ \ $26.91$\ \                         & \ \ $28.02$ \ \                      & \ \ $26.91$\ \                      & \ \ $28.03$\ \                       & \ \ $26.90$\ \ \\
   \hline \hline
 \end{tabular}
 }\label{tab2}}
\end{table}
%=======================================================================================================

In the following, we will show the constraining progress and results
for three dark energy parametrization models: CPL model, JBP model
and Alam model.

%==================== Fig 2.1 =============================================
\begin{figure}[ph]
 \centering
\includegraphics[angle=0,width=0.65\textwidth]{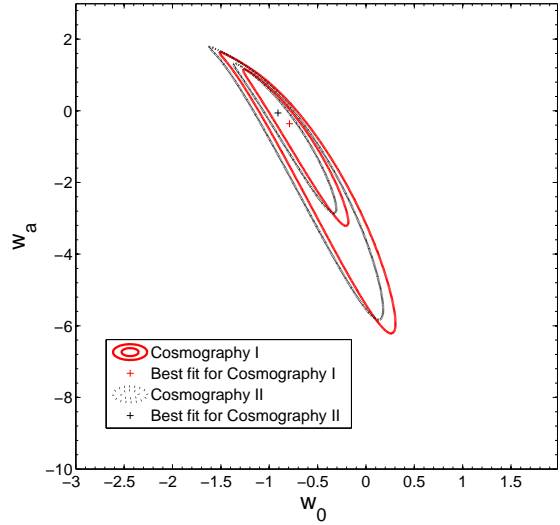}
       \caption{Contours of likelihood in the ($w_0, w_a$) plane
in the CPL dark energy model for a flat universe, from CMB and BAO
observations together with the 42 GRBs data ($z>1.4$) obtained by
the cosmographic method. The red line is for Cosmography I, and the
red plus sign denotes the best-fit values ($w_0=-0.79, w_a=-0.36$).
The black line is for Cosmography II, and the black plus sign
denotes best-fit values ($w_0=-0.91, w_a=-0.06$). The contours
correspond to 1 and 2-$\sigma$ confidence regions.}
           \label{Fig2}
            \end{figure}
%=========================================================================

(1) Chevallier-Polarski-Linder (CPL)
model,\cite{Chevallier2001,Linder2003} in which dark energy with a
parametrization EoS:
\begin{equation}
w(z)=w_0+w_a\frac{z}{1+z}.\\
\end{equation}
The corresponding luminosity distance for a flat universe is
\begin{equation}
d_L=cH_0^{-1}(1+z)\int_0^z{\frac{dz}{E(z)}},\\
\end{equation}
where
\begin{equation}
E(z)=\big[(1+z)^3\Omega_M+(1-\Omega_M)(1+z)^{
3(1+w_0+w_a)}e^{-3w_az/(1+z)}\big]^{-1/2}.\\
\end{equation}

We set $H_0$ and $\Omega_M$ as free parameters and their prior
values are $64 \textrm{km s}^{-1} \textrm{Mpc}^{-1} \leq H_0 \leq 80
\textrm{km s}^{-1} \textrm{Mpc}^{-1}$ and $0.23 \leq \Omega_M \leq
0.31$ respectively. We also assume this prior in the following
analysis. For cosmography I, we find $\chi_{\textrm{min}}^2=28.03$
at \{$H_0$ , $\Omega_M$\}=\{$80 \textrm{km s}^{-1}
\textrm{Mpc}^{-1}$, 0.29\}, the best fitting values are
$w_0=-0.79^{+0.38}_{-0.34}$ and $w_a=-0.36^{+1.15}_{-1.7}$
(1$\sigma$). For cosmography II, we find
$\chi_{\textrm{min}}^2=26.91$ at \{$H_0$ , $\Omega_M$\}=\{$75
\textrm{km s}^{-1} \textrm{Mpc}^{-1}$, 0.28\}, the best fitting
values are $w_0=-0.91^{+0.38}_{-0.32}$ and
$w_a=-0.06^{+1.06}_{-1.66}$ (1$\sigma$). Fig. 2 shows the
constraints on $w_0$ and $w_a$ parameters for the CPL model. We
present the best-fit value of $w_0$ and $w_a$ with $1\sigma$
uncertainties,   as well as the best-fit values of $\Omega_M$,
$H_0$, and $\chi_{\rm min}^2$, for the CPL model in Table 2.

(2) Recently, Jassal, Bagla \& Padmanabhan argued that in CPL model
some problems will present at high redshifts, they thus proposed a
new parametrization EoS (JBP model):\cite{Jassal2004}
\begin{equation}
w(z)=w_0+w_a\frac{z}{(1+z)^2},\\
\end{equation}
which can model a dark energy component that has the same value at
lower and higher redshifts, with rapid variation at low \textit{z}.
The corresponding luminosity distance  for a flat universe is the
same with Eq.(26), and Eq.(27)  becomes

\begin{equation}
E(z)=\big[(1+z)^3\Omega_M+(1-\Omega_M)(1+z)^{
3(1+w_0)}e^{3w_az^2/2(1+z)^2}\big]^{1/2}.\\
\end{equation}

For cosmography I, we find $\chi_{\textrm{min}}^2=28.02$ at \{$H_0$
, $\Omega_M$\}=\{$80 \textrm{km s}^{-1} \textrm{Mpc}^{-1}$, 0.29\},
the best fitting values are $w_0=-0.74^{+0.53}_{-0.51}$ and
$w_a=-0.93^{+3.17}_{-3.53}$ (1$\sigma$). For cosmography II, we find
$\chi_{\textrm{min}}^2=26.91$ at \{$H_0$ , $\Omega_M$\}=\{$79
\textrm{km s}^{-1} \textrm{Mpc}^{-1}$, 0.25\}, the best fitting
values are $w_0=-1.37^{+0.52}_{-0.51}$ and
$w_a=2.02^{+3.16}_{-3.48}$ (1$\sigma$). Fig. 3 shows the constraints
on $w_0$ and $w_a$ parameters for the JBP model. We present the
best-fit value of $w_0$ and $w_a$ with $1\sigma$ uncertainties,   as
well as the best-fit values of $\Omega_M$, $H_0$, and $\chi_{\rm
min}^2$, for the JBP model in Table 2.

%==================== Fig 2.2 ==========================================================================
\begin{figure}[pb]
 \centering
\includegraphics[angle=0,width=0.65\textwidth]{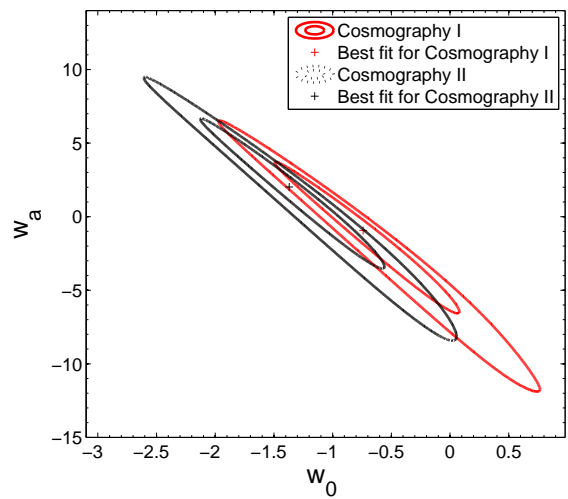}
       \caption{Same as Fig. 2, but  fitting the JBP model. For Cosmography I: the best-fit values ($w_0=-0.74,
w_a=-0.93$); for Cosmography II: the best-fit values ($w_0=-1.37,
w_a=2.02$).}
           \label{Fig3}
            \end{figure}

(3) The third dark energy parametrization model that we consider
is\cite{Alam2003}
\begin{equation}
w(z)=\frac{1+z}{3}\frac{A_1+2A_2(1+z)}{\Omega_{\rm DE}(z)}-1,\\
\end{equation}
where
\begin{equation}
\Omega_{\rm DE}(z)=A_1(1+z)+A_2(1+z)^2+1-\Omega_M-A_1-A_2,\\
\end{equation}
thus
\begin{equation}
E(z)=\big[(1+z)^3\Omega_M+A_1(1+z)+A_2(1+z)^2+1-\Omega_M-A_1-A_2\big]^{1/2}.\\
\end{equation}

For cosmography I, we find $\chi_{\textrm{min}}^2=28.03$ at \{$H_0$
, $\Omega_M$\}=\{$79 \textrm{km s}^{-1} \textrm{Mpc}^{-1}$, 0.29\},
the best fitting values are $A_1=0.6^{+0.96}_{-0.97}$ and
$A_2=-0.11^{+0.26}_{-0.25}$ (1$\sigma$). For cosmography II, we find
$\chi_{\textrm{min}}^2=26.90$ at \{$H_0$ , $\Omega_M$\}=\{$78
\textrm{km s}^{-1} \textrm{Mpc}^{-1}$, 0.26\}, the best fitting
values are $A_1=-0.39^{+0.9}_{-0.87}$ and
$A_2=0.1^{+0.24}_{-0.24}$(1$\sigma$). Fig. 4 shows the constraints
on $A_1$ and $A_2$ parameters for the Alam model. We present the
best-fit value of $A_1$ and $A_2$ with $1\sigma$ uncertainties,  as
well as the best-fit values of $\Omega_M$, $H_0$, and $\chi_{\rm
min}^2$, for the Alam model in Table 2.

We can find that when combining the GRBs dataset obtained by
cosmographic fitting method with the CMB and BAO observations, the
combined data can give more stringent results on different dark
energy parametrization models. It is noted that all these results
for above models are consistent with the $\Lambda$CDM model in
1-$\sigma$ confidence region.

%==================== Fig 2.3 ==========================================================================
\begin{figure}[pb]
 \centering
\includegraphics[angle=0,width=0.65\textwidth]{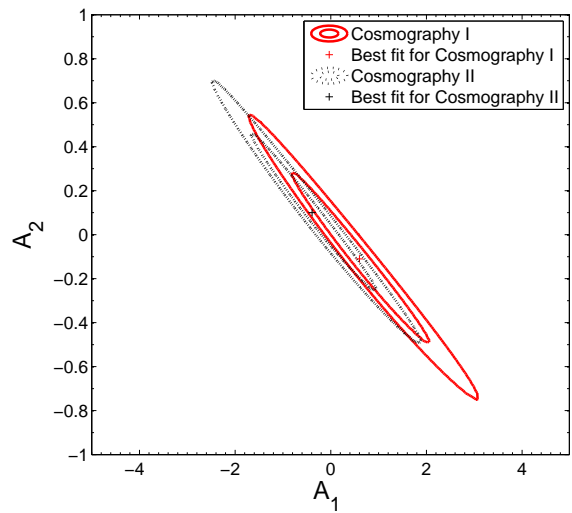}
       \caption{Same as Fig. 2, but in the ($A_1, A_2$) plane
in the Alam model for a flat universe. For Cosmography I: the
best-fit values ($A_1=0.6, A_2=-0.11$); for Cosmography II: the
best-fit values ($A_1=-0.39, A_2=0.1$).}
           \label{Fig4}
            \end{figure}

\section{Summary and Discussion}
Due to the lack of the GRB sample at low redshift, there has been a
so-called circularity problem which can  always be a obstacle for
applying GRBs data to constrain cosmological parameters. Based on
the basic assumption that objects at the same redshift should have
the same luminosity distance, Liang \textit{et al.} proposed a new
method to calibrate GRB relations in a completely
cosmology-independent way, namely obtaining the distance modulus of
a GRB by interpolating from the Hubble diagram of SNe Ia and then
calibrate the GRB relations with these calculated distance
moduli\cite{Liang2008a}. There is a well-known fitting formula in
cosmography, which can reflect the Hubble relation between
luminosity distance and redshift with cosmographic parameters which
can be fitted from SNe Ia. In this work,  we propose another
approach to calibrate GRB luminosity relations with cosmography
fitting from SNe Ia data.  We adopted the fitting results from  the
Union set of  SNe Ia for the so-called Cosmography I and
II,\cite{Vitagliano2010} and calibrate five GRB relations by this
cosmographic fitting method. The calibration results obtained using
two cosmographic fitting are fully consistent with each other.
Assuming that GRB luminosity relations do not evolve with redshift,
we obtained the distance modulus of the GRB data at higher redshift
$1.4<z\le6.6$. Note that the circularity problem could be completely
avoided if we apply these GRBs data to constrain cosmological
parameters. We thus constrained three parameterization dark energy
models of the CPL, JBP and Alam models by combining the GRB data at
high redshift with the CMB and BAO observations, and the constraint
results are all consistent with the $\Lambda$CDM model in 1-$\sigma$
confidence region.

%\clearpage

Compared to previous cosmology-independent calibration method, we
find that results obtained by the cosmographic method consistent
with those calibrated by using the interpolation method and the
iterative procedure from SNe Ia with the same GRB
sample\cite{Liang2008a,Liang2008b}; this situation are consistent
and similar with the recent conclusions that the cosmographic Amati
relation agrees in the errors with other cosmology-independent
calibrations,\cite{Capozziello2010}  and that the calibration
coefficients and the intrinsic scatter actually do not depend on the
adopted calibration procedure seriously, based on the use of a
fiducial model, the cosmographic method and the local
regression.\cite{Cardone2011} It should be noted that the fitting
procedure used in Kodama \textit{et al.} \cite{Kodama2008} depends
seriously on the choice of the formula. In Cosmography, the
cosmographic formula is totally independent of any cosmological
models and could accurately evaluate the Hubble relation between
luminosity distance and redshift, and can be considered as a perfect
fitting function to calibrate the GRB relations using SNe Ia data.
The reliability of this method should be more reasonable than the
fitting procedure which chooses a formula in
arbitrary.\cite{Kodama2008} However, since the distance moduli of
GRBs calculated by the fitting method including the cosmographic
method are obviously relevant to all of SN Ia data points for
calibration, it will be not appropriate for directly combining these
two datasets to constrain the cosmological
parameters.\cite{Kodama2008,Tsutsui2009} But like showing in this
work, we can combine the GRBs dataset obtained by cosmographic
fitting method with the CMB and BAO observations and the combined
data can give more stringent results. As to the interpolation
method,\cite{Liang2008a} for deducing one individual GRB, only a few
SNe Ia close to this GRB were used. In this case, the measurement
error for a single SNe Ia's modulus may influence the final result
of calibration. In the cosmographic method, the full information of
SNe Ia data is completely used in the calibration. It should be
noted that our approach provides no more accurate cosmological
parameter constraints than other works such as the simultaneous fit
method.\cite{Basilakos2008} However, the primary motivation of this
work is not on improvement of the statistical error, rather on
avoiding the circularity problem more clearly in logic.

For comparing GRB to SNe Ia, GRBs are almost immune to dust
extinction, whereas SNe Ia observations suffer extinction from the
interstellar medium. GRBs can extend the Hubble diagram to much
higher redshifts beyond SNe Ia data. Different dark energy models
may have very different Hubble diagrams at high redshifts. On the
other hand, due to the large statistical scatters of the relations
and the small dataset compared to SNe Ia, the contribution of GRBs
to the cosmological constraints would not be sufficiently
significant at present. However, it should be noted that a single
GRB at high redshift will provide more information than a single
maximal redshift SNe Ia,\cite{Schaefer2007} and the larger scatter
of GRB data is somewhat coming from the observational limit of
nowadays technological level, which means the GRB data may
eventually turn into good data as our observational capability
enhance. It has been found that cosmological constraints would
improve substantially with more simulated GRBs expected by future
observations through Monte Carlo simulations.\cite{Bertolami2006}
With more and more GRBs observed from Fermi Gamma-ray Space
Telescope with much smaller scatters, and its combination with the
increasing Swift data, GRBs could be used as an additional choice to
set tighter constraints on cosmological parameters of dark energy
models. At that time when GRB data becomes common used data sample
to constrain high redshift cosmology, the so called circularity
problem for GRB data will be a key problem to solve. Moreover, it is
noted that accurately calibrating the GRB relations is also very
important for the GRB theory system itself even without considering
the cosmology. We have to note  stress again that the high-redshift
GRB dataset we obtained here is completely cosmology-independent, it
will ultimately fill the data crack between SNe Ia and CMB after
more GRBs being observed in the future.

\section*{Acknowledgments}

%We thank  Yun Chen, Shuo Cao, Hao Wang, Fang Huang, Jie Ma,
%Xingjiang Zhu, and Dr. Yi Zhang for discussions.
This work was supported by the National Natural Science Foundation
of China under the Distinguished Young Scholar Grant 10825313 and
Grant 11073005, the Ministry of Science and Technology national
basic science Program (Project 973) under Grant No.2007CB815401, and
"the Fundamental Research Funds for the Central Universities".

%\clearpage

%\begin{thebibliography}{000} %for 3 digits
%\begin{thebibliography}{00}  %for 2 digits

\end{document}